\documentclass[aps,onecolumn]{revtex4}
\newcommand{\be}{\begin{equation}}
\newcommand{\ee}{\end{equation}}
\newcommand{\bear}{\begin{eqnarray}}
\newcommand{\eear}{\end{eqnarray}}

\begin{document}

\title{Mass dependence of the vacuum energy density in the \\
massive Schwinger model}
\author{Taekoon Lee}
\email[]{tlee@kunsan.ac.kr}
\affiliation{Department of Physics, Kunsan National University,
Kunsan 573-701, Korea}

\begin{abstract}
The vacuum energy density of the massive Schwinger model is shown
to be not power expandable in the fermion mass.
\end{abstract}
\maketitle

The exact operator solution of the massless Schwinger model \cite{schwinger}
shows that the Hilbert space of the Schwinger model 
is given as the product of that of a free, massive scalar field 
and the theta vacua \cite{ls,rw,morchio}. The theta vacua are 
known to arise from the
anomalous breakdown of the chiral symmetry and the nontrivial
topology of  the field configurations in two-dimensional $U(1)$
gauge theory. The vacua may be labeled by the phase of the
chiral condensate.

Although the physics of the Schwinger model should not depend on 
any particular choice of the vacuum among the theta vacua
one could ask whether it is at all possible  to choose a vacuum
in the Schwinger model.
In a theory with the ordinary spontaneous symmetry breaking
a particular vacuum may be picked up out of the continuum of vacua
by adding an infinitesimal symmetry breaking term 
in such a way that the energy shift due to the
added term be minimized at the selected vacuum. This is the
vacuum alignment \cite{dashen}. 

In the case of the theta vacua it was argued
that the vacuum alignment is not possible owing to the absence of
the Goldstone boson associated with the anomalous chiral
symmetry breaking \cite{cjs}. The implication of this is that
the phase of the chiral condensate in the Schwinger model
cannot be picked
up by adding an infinitesimal fermion mass term to 
the Schwinger model.
Instead, it is widely accepted that a theta vacuum, and the
corresponding
phase in the chiral condensate, can be selected
by adding a topological term  to the Schwinger
model. This is peculiar, however, since it would suggest that
 the original Schwinger model have a unique vacuum 
which, if true, would  contradict  the operator solution.
 
A consequence of the above uniqueness of the vacuum in the Schwinger
model is the power expandability of the vacuum energy of the
massive Schwinger model in the fermion mass \cite{cjs,leut.smil}.
To be
specific, consider the Lagrangian density 
of the massive Schwinger model
 \bear
{\cal L}_{\rm A}&=& -\frac{1}{4} F_{\mu\nu} F^{\mu\nu} +i 
 \bar{\psi}
 (\not\!\partial+i e \not\!\! A)
 \psi +|m|( e^{i\alpha }\bar{\psi}_{\rm L} \psi_{\rm R} +
 \text{h.c.})\,,
\label{lag1}
\eear
 where  $\psi_{\rm L,R}=\frac12 (1\pm\gamma_5)\psi$ and
 $\alpha$ denotes the phase of the fermion mass $m=|m|e^{i\alpha}$.

Usually,  invoking parity symmetry, the vacuum of the 
massless Schwinger model is assumed to have a real chiral condensate,
which allows one to write the
chiral  condensate in the vacuum of the model
(\ref{lag1}) as
 \bear
<\!\bar\psi_{\rm L} \psi_{\rm R}\!>_{\rm A}=\Sigma +O(m),
\label{con_A}
\eear
where $\Sigma$ is a real parameter and $<\,\,>_{\rm A}$ denotes
the expectation values in the vacuum of  (\ref{lag1}).
Then using the mass perturbation
the vacuum energy density of (\ref{lag1}) can be written as
\bear
\varepsilon_{\rm A}=\varepsilon_0 -
 \Sigma (m +m^*) +O(m^2)\,,
\label{energy1}
\eear
 where
$\varepsilon_0$ denotes the vacuum energy density of the
massless Schwinger model.  The vacuum energy density thus
can be expanded in powers of the fermion mass
and its complex conjugate.

On the other hand, if the theta vacua are alignable by the
fermion mass term, the chiral condensate
would get a phase that cancels  the fermion mass phase exactly
in order to minimize the energy shift by the mass term.
The vacuum energy density then
would depend on $|m|$ only, and it could not be expanded in
powers of the fermion mass. Therefore, whether the theta vacua
are alignable or not can be determined by looking at the mass
dependence of the vacuum energy density.

In this note we give a simple argument that shows the vacuum
energy density (\ref{energy1}) cannot be compatible with the
anomaly equation for the chiral symmetry,
and show that the vacuum energy density that is consistent
with the anomaly equation depends only
on the magnitude $|m|$ of the fermion mass.

To show this we consider the following  chirally rotated form of the
Lagrangian density (\ref{lag1}) \cite{fugi}
\bear
{\cal L}_{\rm B}&=& -\frac{1}{4} F_{\mu\nu} F^{\mu\nu} +i 
 \bar{\psi}(\not\!\partial+i e \not\!\! A) \psi 
 +|m|( e^{i(\alpha -\beta)}\bar{\psi}_{\rm L} 
 \psi_{\rm R} +\text{h.c.}) +
\frac{ \beta}{2} \tilde F,
\label{lag2}
\eear
where $\beta$ is a real
parameter and $\tilde F$ denotes the topological term given by
\bear
\tilde F=\frac{e}{2\pi}\epsilon^{\mu\nu}F_{\mu\nu}\,,
\eear
where $\epsilon^{\mu\nu}$ is an antisymmetric tensor 
with $\epsilon^{01}=1$.
The difference ${\cal \delta H}$ of the Hamiltonian densities of the
model (\ref{lag2}) and the massless Schwinger model is given,
up to an irrelevant, field-independent constant term, by
\bear
{\cal\delta H}=-|m|(e^{i(\alpha -\beta)}\bar{\psi}_{\rm L}\psi_{\rm R}
+\text{h.c.})-\frac{\beta}{2} \tilde F\,,
\eear
and the vacuum energy density of  (\ref{lag2}) then by
\bear
\varepsilon_{\rm B}=\varepsilon_0 - 2 |m| {\rm Re}(e^{i(\alpha-\beta)}
<\!\bar\psi_{\rm L}\psi_{\rm R}\!>_{\rm B})-\frac{\beta}{2}
<\!\tilde F\!>_{\rm B}+O(m^2)\,,
\label{energy2}
\eear
where  $<\,\,>_{\rm B}$ denotes the
expectation values in the vacuum of (\ref{lag2}).

Now suppose the condensate (\ref{con_A}) is correct. Then the chiral
condensate of the model (\ref{lag2}) will be given as
\bear
<\!\bar\psi_{\rm L}\psi_{\rm R}\!>_{\rm B}=\Sigma e^{i \beta} +O(m)\,,
\label{con_B}
\eear
 since the  condensates of the two vacua
 are related by $<\!\bar\psi_{\rm L}\psi_{\rm
R}\!>_{\rm A}=<\!\bar\psi_{\rm L}\psi_{\rm R}\!>_{\rm B}e^{-i \beta}$.
Using the anomaly equation for the chiral current
$J^\mu_5=\bar\psi\gamma^\mu\gamma_5\psi$
the expectation value of the topological charge density
can be related to the chiral condensate by
\bear
0=\partial \mu <\!J^\mu_5\!>_{\rm B}=  2i |m| \big(e^{i (\alpha-\beta)}
<\!\bar\psi_{\rm L} \psi_{\rm R}\!>_{\rm B} -e^{-i (\alpha-\beta)}
<\!\bar\psi_{\rm R} \psi_{\rm L}\!>_{\rm B}\big) -
<\!\tilde F\!>_{\rm B}\,,
\label{anomalyeq}
\eear
which  gives
\be
 <\!\tilde F\!>_{\rm B}=-4|m|\Sigma \sin\alpha +O(m^2)\,.
\label{con_topo}
\ee
Substituting (\ref{con_B}) and (\ref{con_topo}) into (\ref{energy2})
we get
\bear
\varepsilon_{\rm B}=\varepsilon_0 - 2 |m|\Sigma\cos\alpha+2
|m|\Sigma\beta\sin\alpha +O(m^2)\,.
\label{energy2A}
\eear
Since the Lagrangian densities (\ref{lag1}) and (\ref{lag2}) are
equivalent
$\varepsilon_{\rm A}$ and $\varepsilon_{\rm B}$ must have the same $m$
dependence for 
any $\beta$ but
obviously this is impossible with (\ref{energy2A}). We, therefore,
conclude that
the widely accepted expressions (\ref{con_A}) and (\ref{energy1})
 cannot be compatible with the anomaly equation.

To find the chiral condensate that is consistent with the anomaly
equation we  modify the condensate (\ref{con_A}) by 
adding a phase factor
as
\bear
<\!\bar\psi_{\rm L} \psi_{\rm R}\!>_{\rm A}=\Sigma e^{i\chi} +O(m)\,,
\label{con_A2}
\eear
where $\Sigma$ is now assumed to be a positive real parameter.
The chiral condensate of (\ref{lag2}) corresponding to this is then
given by
\bear
<\!\bar\psi_{\rm L}\psi_{\rm R}\!>_{\rm B}=\Sigma e^{i (\chi+\beta)}
+O(m)\,,
\label{con_B2}
\eear
and the expectation value for the topological charge density by
\bear
 <\!\tilde F\!>_{\rm B}=-4|m|\Sigma \sin(\alpha+\chi)\,.
\label{con_topo2}
\eear

The vacuum energy density of (\ref{lag1})  will then be given by
\bear
\varepsilon_{\rm A}=\varepsilon_0 -
2 |m| \Sigma \cos(\alpha+\chi) +O(m^2)\,,
\label{energy1A}
\eear
and that of (\ref{lag2}) by
\bear
\varepsilon_{\rm B}=\varepsilon_0 - 2 |m|\Sigma\cos(\alpha+\chi)+2
|m|\Sigma\beta\sin(\alpha+\chi) +O(m^2)\,.
\label{energy2B}
\eear
For these vacuum energy densities to have the identical $m$ dependence
for all
$\beta$ the phase $\chi$  must
satisfy
\bear
\sin(\alpha+\chi)=0 \,,
\label{con_phase}
\eear
which shows the phase of the chiral condensate
is governed by that of the fermion mass.
This means that the theta vacua must be alignable by the
fermion mass term!

Note that Eq. (\ref{con_phase}) shows the vacuum expectation value
for the topological charge
density always vanishes
and the vacuum energy density does not depend  on
the phase of the fermion mass. The implication of this is that
the true vacuum of the massive Schwinger model
is CP symmetric irrespective of the phase of the fermion mass or
the presence of the topological term.
Since the energy shift by the
mass term is minimized at
$\chi=-\alpha$  the vacuum energy density is given by
\bear
\varepsilon_{\rm A}=\varepsilon_{\rm B}=
\varepsilon_0- 2|m \Sigma| +O(m^2)\,.
\label{vacuumenergy2}
\eear

Finally, we note that the argument so far for  the theta vacuum
alignment is applicable to QCD as well, 
and that  the theta vacuum alignment in QCD removes, entirely,
the strong CP phase  of the quark mass matrix 
from the QCD low energy  physics \cite{lee.t}.

\begin{acknowledgments}
This work was supported by the  Korea Research 
Foundation Grant (KRF-2006-015-C00251).
\end{acknowledgments}

\bibliography{vm}

\end{document}